\documentclass[aps,prb,showpacs,twocolumn,preprintnumbers,amsmath,amssymb]{revtex4}
\usepackage{dcolumn}
\usepackage{bm}
\usepackage{amsmath}
\usepackage{feynmp}

\usepackage{graphicx}

\begin{document}

\title{Bulk effects on topological conduction on the surface of 3-D topological insulators}

\author{Quansheng Wu}
\affiliation{Institute of Applied Physics and Computational Mathematics, Beijing 100094, P.R. China}
\affiliation{CAEP Software Center for High Performance Numerical Simulation Beijing, 100088, China}

\author{Vincent E. Sacksteder IV$^{1}$}
\affiliation{Institute of Physics, Chinese Academy of Sciences, Beijing 100190, China}
\affiliation{Division of Physics and Applied Physics, Nanyang Technological University, 21 Nanyang Link, Singapore 637371}




\date{\today}

\begin{abstract}
The surface states of a topological insulator in a fine-tuned magnetic field are  ideal candidates for realizing a  topological metal which  is protected against disorder.  Its signatures are (1) a conductance plateau in long wires in a finely tuned longitudinal magnetic field and (2) a conductivity which always increases with sample size, and both are independent of  disorder strength.   We numerically study how these experimental transport signatures are affected by  bulk physics in the interior of the topological insulator sample.  We show that both signatures of  the topological metal are robust against bulk effects.  However   the bulk does substantially accelerate  the  metal's decay in a magnetic field and alter its response to surface disorder.   When the disorder strength is tuned to resonance with the bulk band  the conductivity follows the predictions of scaling theory, indicating that conduction is diffusive.  At other disorder strengths the bulk reduces the effects of surface disorder and scaling theory is systematically violated, signaling that conduction is not fully diffusive.  These effects  will change the magnitude of the surface conductivity and the  magnetoconductivity.
 \end{abstract}

\maketitle
\section{Introduction}
The Dirac fermions residing on the surface of strong topological insulators (TIs)  provide a new opportunity for realizing a topological metal which remains conducting regardless of disorder strength or sample size.   \cite{Kane05, Min06, Zhang09, Hasan10, Li12, Culcer12, Bardarson13} This metal should be general to any single species of  Dirac fermions which breaks spin  symmetry but retains time reversal symmetry. 
When the Fermi level is tuned inside the TI's bulk band gap,  all states inside the TI bulk are localized.  Conduction can occur only on the TI surface, which hosts   a single species of two-dimensional (2-D) Dirac fermions. \footnote{More generally, any odd number of fermions is possible.}   These  are predicted to remain always conducting regardless of disorder strength, forming a topological metal.  

Many beautiful TI experiments have visualized the Dirac cone, spin-momentum locking, Landau levels with $\sqrt{n}$ spacing, and SdH oscillations. \cite{Hsieh08,Hanaguri10} These signals are visible only when either momentum or the Landau level index are approximately conserved;  they disappear when disorder is strong.  In contrast, the topological metal's hallmark is robust surface conduction at any disorder strength and  in large samples, even when disorder destroys the Dirac cone.  In this article we will focus on experimental signatures of this protection against disorder.

We will show that bulk physics in the TI's interior substantially modifies   the topological metal.  Even though it is a surface state,  in response to disorder it may explore the TI bulk or even tunnel between surfaces.    We used several months on a large parallel supercomputer to perform extensive calculations of conduction, including both bulk and surface physics.  

Two experimental signatures are available for  proving incontrovertibly that a topological metal is indeed robust against disorder.        The  first  is a quantized conductance in   long wires.  In ordinary wires the conductance $G$ decreases with wire length until  only one channel remains open, i.e. $G = \frac{e^2}{h}$, and then transits into a localized phase where the last channel decays exponentially.     In contrast the topological metal exhibits  one \textit{perfectly conducting channel} (PCC) which remains forever topologically protected, so in long wires the conductance is quantized at $G = \frac{e^2}{h}$.   \cite{Zirnbauer92,Mirlin94,Ando02,Takane04,Ryu07,Ostrovsky10,Zhang10,Hong13}  The PCC can be realized only when there is no gap in the surface states' Dirac cone.    In  TI wires locking between spin and momentum creates a gap,  but a specially tuned longitudinal magnetic field   $B$  can be used to close the gap and realize the PCC's quantized conductance.  In this fine-tuned scenario $B$ breaks time reversal symmetry and causes the PCC to eventually decay.       Our results   will confirm the PCC's robustness against bulk effects,
and show that both the optimal value of $B$ and the PCC's optimal decay length are altered by bulk physics.   These results, and the PCC's response to sample dimensions, magnetic fields, and disorder,  will be useful to experimental PCC hunters.
  
   The second key signature of the topological metal concerns the  diffusive regime, where the sample aspect ratio $L/W$ is  small enough that several conducting channels remain open, but the sample is still bigger than the scattering length.  ($W$ and $L$ are the sample width and length.) 
   In this regime, regardless of disorder strength, a topological metal's \textit{ conductivity} $\sigma = G L / W$ \textit{always increases when the length $L$ and width $W$ are increased in proportion to each other.}    \cite{Ostrovsky07,Nomura07,Bardarson07,SanJose07,Tworzydlo08,Lewenkopf08,Adam09,Mucciolo10,Mong12}  This signature contrasts with non-topological materials  where the conductivity can always be forced to decrease by making the disorder large enough.  \cite{Asada04,Nomura07}   In both cases the increasing conductivity - called weak antilocalization (WAL) - can be removed by introducing a weak magnetic field; experimental studies of the  TI magnetoconductivity  are extremely popular.   In the diffusive regime both the conductivity and the magnetoconductivity are expected to follow universal curves prescribed by scaling theory, independent of  sample  details.
   
    Here we will  confirm that the always-increasing conductivity is robust against both bulk effects and very strong disorder.  We will also show that the bulk reduces scattering and causes violation of the universality predicted by scaling theory, which implies that the topological metal's conduction is not purely diffusive.  This has immediate consequences for experiments: we expect that magnetoconductivity measurements are sensitive to bulk physics, and that the magnitude of the observed signal will vary systematically with the Fermi level and disorder strength.
    
    Lastly, we find that the bulk reduces the effects of scatterers residing on the TI surface.  The topological metal is free to reroute around scatterers, into the TI bulk.  This is a second level of topological protection, in addition to the well-known suppression of backscattering.  It should change the  surface conductivity and increase the topological metal's robustness  against issues of sample purity, substrates, gating,  etc.

      \section{The Model}
      The topological metal is independent of any short-scale variation in the TI sample, including any microscopic details of the Hamiltonian. Its two signatures are regulated by only  two parameters: the scattering length and localization length.  Because we are concerned with Fermi energies inside the bulk band gap, bulk effects on the topological metal  can depend on  only a small number of parameters - the Fermi level, the Fermi velocity,  the penetration depth, the bulk band gap, and the bulk spectral width.  Therefore we study a  computationally efficient  minimal tight binding model of a strong TI implemented on a cubic lattice with dimensions $H,W$, and $L$.     With four orbitals per site, the model's momentum representation is:
\begin{eqnarray}
\mathcal{H}(\vec{k}) & = & 2  \Gamma^1  - \frac{1}{2} \sum_{i=1}^3 (\Gamma^1 - \imath \Gamma^{i+1}) e^{-\imath k_i a}  + H.c.  
\end{eqnarray}
\noindent $\Gamma^i$ are the Dirac matrices $1 \otimes \sigma_z, -\sigma_y \otimes \sigma_x, \sigma_x \otimes \sigma_x, -1 \otimes \sigma_y$,  $a=1$ is the lattice spacing, and the penetration depth is $d \propto a$.   \cite{Schubert12} We include a magnetic field oriented longitudinally along the axis of conduction by multiplying the hopping terms by Peierls phases.  \footnote{We neglect the Zeeman term.  This term will be small and proportional to $1/W^2$ because the PCC conductance plateau occurs at a value of $B$ which is proportional to $1/W^2$.}  This non-interacting model exhibits a spectral width $\Delta E = 10$,  a bulk band gap in the interval $E =[-1,1]$,  a single Dirac cone in the bulk gap, and Fermi velocity $v_F = 2$.     To this model  we add uncorrelated white noise  disorder $u(x)$.  On each individual site the disorder is proportional to the identity and its strength is chosen randomly from the interval $\left[ -U/2, \, U/2 \right] $, where $U$ is the disorder strength.   
   We attach  two clean semi-infinite leads that have $W \times H$  cross-sections equal to that of the sample itself and  evaluate the conductance using the Caroli formula\cite{Caroli71,Meir92} $G = -\frac{e^2}{h}{Tr}((\Sigma^r_{L}-\Sigma^a_{L}) G^r_{LR} (\Sigma^r_{R}-\Sigma^a_{R}) G^a_{RL})$.   $G^a, G^r = (E_F - \mathcal{H} - u \mp \imath \epsilon)^{-1} $ are the advanced and retarded single-particle Green's functions connecting the left and right leads, and $\Sigma_{L,R}$ are the lead self-energies. \cite{Lopez84} 

 In our study we will leave the TI bulk pure, since the main effects of bulk disorder  on the surface states   can be duplicated by adjusting the bulk parameters.  In particular, the bulk band width widens, the bulk band gap narrows, and the penetration depth increases.  These parameters are only weakly sensitive to disorder when the disorder is small compared to the band width. When the disorder strength approaches the band width the bulk states gradually delocalize, surface states are eventually destroyed by tunneling through the bulk, and the material ceases to be a TI.   \cite{Groth09,Guo10,Chen12,Xu12,Kobayashi14}  
 In contrast our focus here is on bulk effects on a healthy topological metal, which are controlled only by the few  bulk parameters which we have listed.
  We include only  surface disorder, which has been extensively investigated experimentally because practical TI devices may be capped, gated, bombarded, or left exposed to atmosphere. \cite{Analytis10, Brahlek11, Aguilar12, Tereshchenko11, Hsieh09,Noh11,Liu12, Lang11, Chen10,Kong11}

\section{The Perfectly Conducting Channel}
 In Figures 1 and 2 we focus on  the topological metal's PCC signature, which is  a quantized conductance plateau seen in very long samples.   We set the disorder strength to  $U =2$, and the Fermi energy is $E_F = 0.7$ in both the sample and the leads.   Figure 1a  shows the plateau in $W \times H$ slabs, which retain all bulk physics but are simplified by avoiding  the gap in the surface Dirac cone.  The plateau conductance is  $G = 2 \frac{e^2}{h}$ because  each surface hosts a PCC.  We define the   PCC decay length $\lambda$, i.e. the plateau length, as the wire length where the conductance is half of its plateau value.     \footnote{The supplementary material shows that in slabs $\lambda$ is roughly independent of the slab width $W$, and that $W=3$ gives results that are quite close to the converged value.  In Figure 1 we use narrow $W=3$ slabs, allowing calculation of very long wires.} 
Figure 1c shows that $\lambda$ grows exponentially as the slab height  $H$ increases from $3$ to $8$, which proves that the PCC decay  is caused by tunneling and is exponentially small except in very thin slabs. \cite{Wu13} This is disorder-enabled tunneling; $\lambda$ diverges in pure slabs.

      \begin{figure}[]
\includegraphics[width=6.2cm,clip,angle=270]{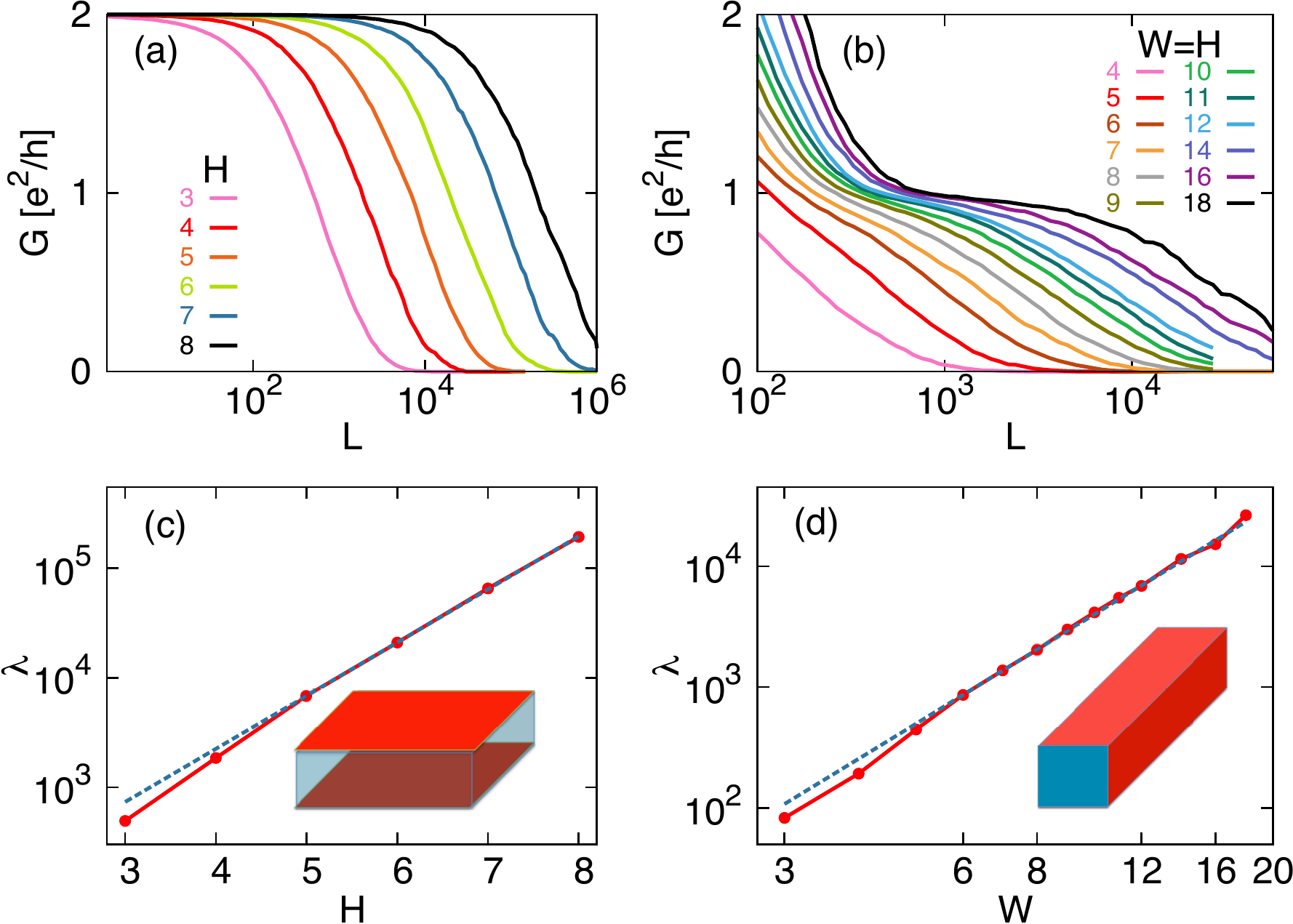}
\caption{ (Color online.)   The perfectly conducting channel (PCC), a topologically protected conductance plateau in very long TI samples.  (a,b) The conductance plateau as a function of  length $L$ in (a) $W \times H$ slabs with no magnetic field, and in (b) $ W\times W$  wires with a fine-tuned longitudinal field.   (c) In slabs the  decay length $\lambda$ is controlled by tunneling and scales exponentially with slab height $H$.       (d) In wires the decay length  scales with $W^3$ and is caused by the surface state's penetration into the bulk.   The dotted lines are (c)  $\lambda \propto \exp(1.12 \; H)$ and  (d)   $\lambda = 4 W^3$.
}
\label{Fig1PCC}
\end{figure}

Next we turn to $W \times W$ TI wires, where the PCC is realized only after we remove the gap in the Dirac cone.   Since spin is locked to  momentum in TIs, parallel transport of the spin on one circuit around a wire's circumference  causes a $\pi$ Berry phase. This causes a  small gap $\Delta_B = 2 \sqrt{2} v_F/C$ where $C$ is the wire circumference.   However the $\pi$ phase can be canceled and the gap removed if the magnetic flux through the wire is fine-tuned to produce an additional $\pi$ phase. \cite{Ostrovsky10,Rosenberg10,Egger10,Zhang10,Bardarson10}  
Figure 1b shows our results after numerically optimizing  $B$  to maximize the PCC lifetime, which at leading order kept the magnetic length $L_B \propto \sqrt{\hbar/eB}$  proportional to the wire width $W$.      The obvious PCC conductance plateau   confirms that the topological metal is robust in wires. 

Figure 1d presents the first numerical calculations of  the PCC decay length $\lambda$'s dependence on the wire width $W$ in a model which includes the TI bulk.  Our results  are of high accuracy, with errors of a few percent.  They  required calculation of very long and wide wires, numerical optimization of $B$, and   many  samples (from $N=864$ for $W \leq 12$ to $N=80$ for $W=18$).   They prove that PCC decay in wires is much faster than a slab's exponentially slow decay, and that $\lambda$ scales  with the cube of the wire width $\lambda \approx \alpha W^3$,   with prefactor $\alpha \approx 4$.  This points to the  magnetic  field and not tunneling as the dominant  source of PCC decay in TI wires.

The decay length's cubic $W^3$ scaling is caused by bulk physics. Previously  $\lambda \propto W^4$ scaling was predicted based on a bulk-independent mechanism \cite{Zhang10, Ando06}.  Diffusion into the bulk  \cite{Altshuler81} is also expected to scale with $L_B^4 \propto W^4$.
 If we note that the large value of $\alpha$ prohibits instances of the  inverse scattering length $ l^{-1} \approx 1/30 a$ (see the supplemental material), then  simple dimensional analysis obtains $\lambda \propto   W^3 /  d^2$, where $d \propto a$ is the penetration depth of the surface states.   This  implies that the fastest mechanism of PCC decay  is caused by the surface state's penetration into the bulk.

           \begin{figure}[]
\includegraphics[ width=3.6cm,angle=270]{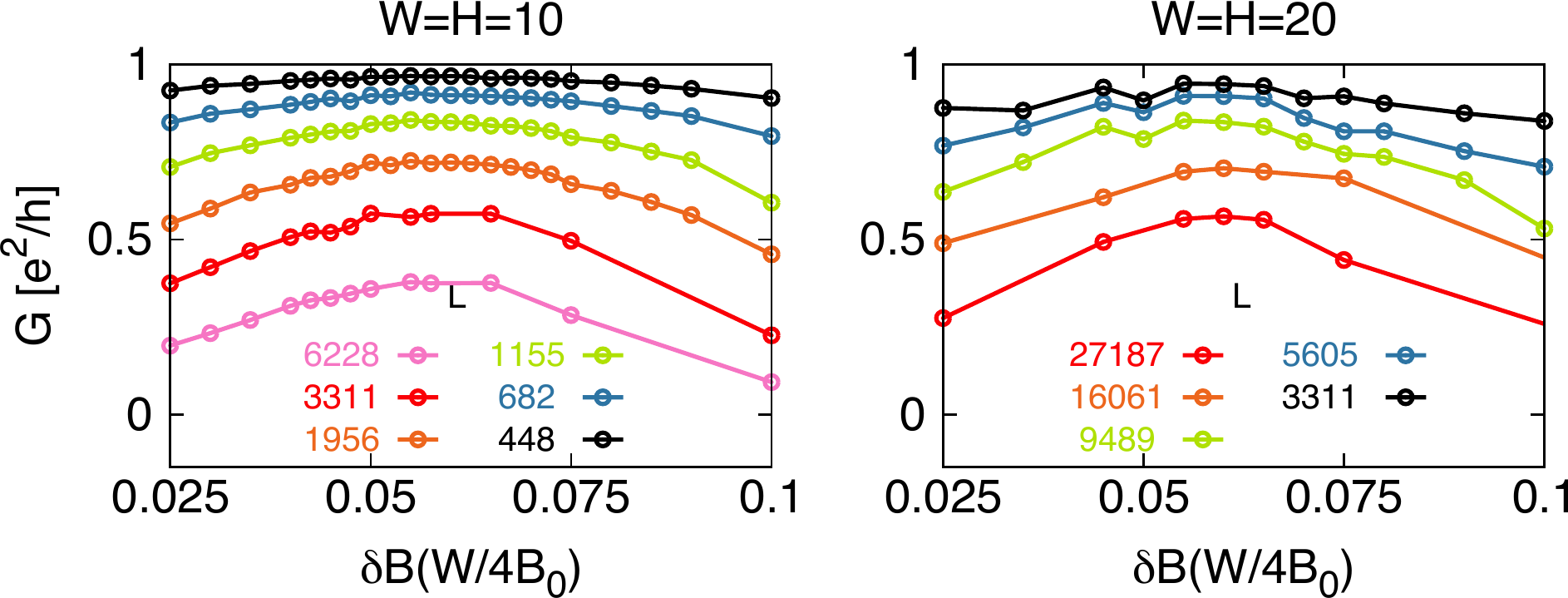}
\caption{ (Color online.) The conductance plateau's dependence on magnetic field $B$ at disorder strength $U=2$.  The wire lengths of the $20 \times 20$ wires (right pane) are chosen to be about $8$ times bigger  than the wire lengths of the $10 \times 10$ wires (left pane).  On the ${x}$ axis  we plot the magnetic field, rescaled by $W/4 B_0 \propto W^3$. The match between the $10 \times 10$  and $20 \times 20$ wires proves that  the decay length scales with $W^3$ and that  the conductance peak's width (in $B$) scales with $W^{-3}$.  At $U=2$ bulk effects cause the conductance peak to shift to a larger field strength, so we have subtracted from $B$  the optimal field strength $B_0$ at zero disorder.
   }
\label{Fig2RShift}
\end{figure}

Bulk physics also combines with surface disorder to alter the magnetic field strength which maximizes $\lambda$.  In any TI disorder on the surface will push the topological states into the bulk,  rescaling their magnetic cross-section by  $(1 -  4 \delta / W)$.  \cite{Schubert12,Wu13,Dufouleur13}  $\delta$ is their displacement, which can be determined from the shift in optimal magnetic field via $\delta = (B_{opt} - B_0) (W/4 B_0)$, where $B_{opt}$ and $B_0$ are respectively the optimal fields at finite disorder $U$ and at zero disorder $U=0$.    Figure 2 plots the conductance at $U=2$ as a function of $ \delta B \, (W/4 B_0), \; \delta B = B - B_0$.  It shows that   the  displacement at $U=2$ is about $0.06$ lattice units and the change in optimal field is $B_{opt} - B_0 = 0.24 \,B_0 / W$.

The two panes of Figure 2 show that  after rescaling by $W/4 B_0$ the conductance peak has the same position and width in both $10 \times 10$ and $20 \times 20$ wires.  This proves that  both the optimal value and the peak width of the  magnetic field  scale with $  B_0 \delta  / W \propto  W^{-3}$.    In thick wires the PCC will not be visible unless the magnetic field is very finely tuned with an accuracy proportional to $W^{-3}$.  We expect that this formula, the previous scaling formulas, and the graphs of the PCC peak will all be useful for PCC hunters.

          \begin{figure}[]
\centering
\includegraphics[angle=90,scale=0.45,clip]{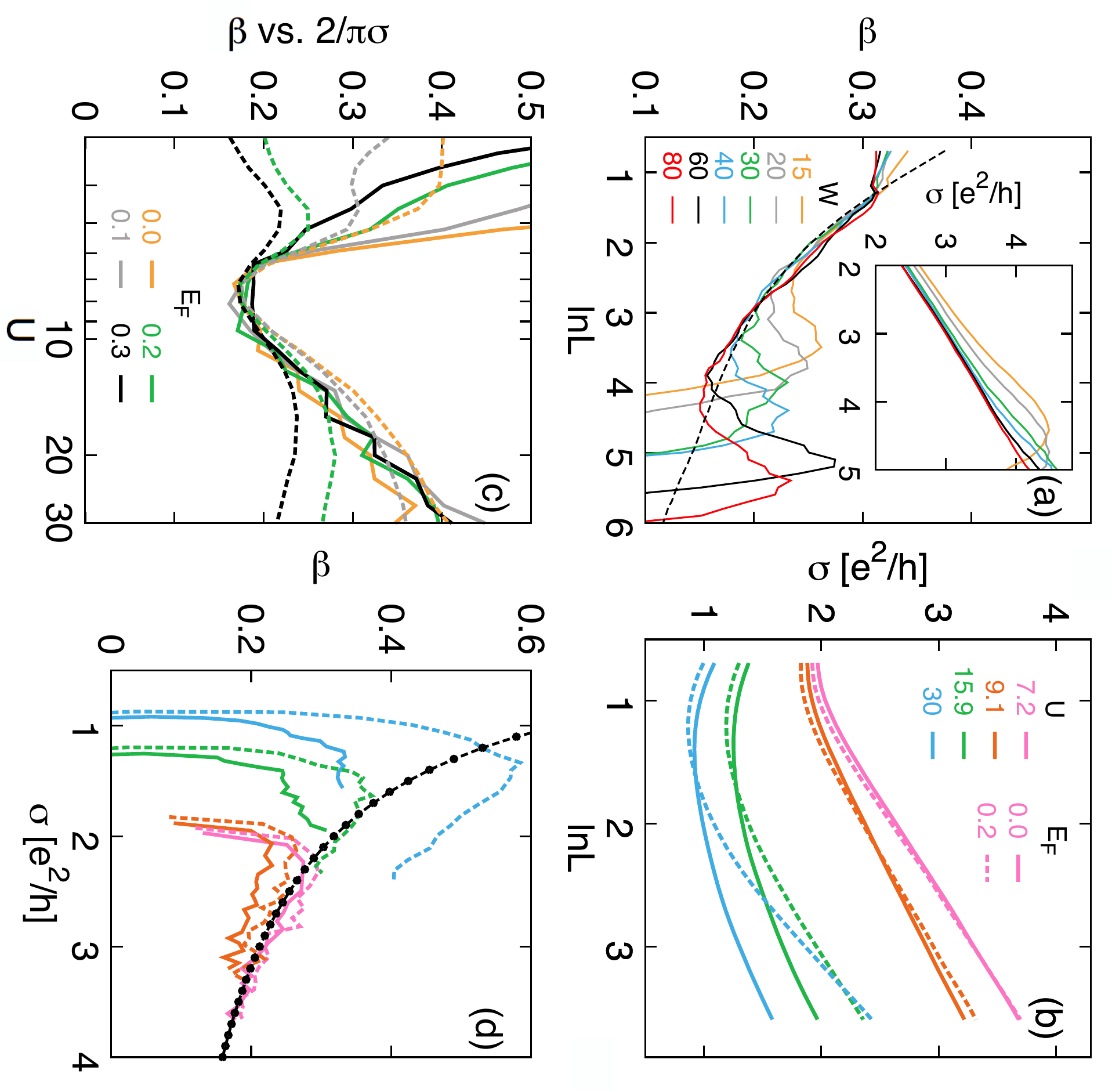}
\caption{ (Color online.) The conductivity $\sigma$ and its logarithmic derivative $\beta(L) = d \ln \sigma / d \ln L $.  (a) Convergence of $\beta$ and $\sigma$ (in inset) as the sample width $W$ is increased from $15$ to $80$.  
   (c) The resonance between the topological surface state and the bulk band, manifested as valleys in both $\beta$ (solid lines) and $2 / \pi \sigma$ (dashed lines).  At the resonance center the topological state is most strongly disordered, and also exhibits the largest conductivity.  (b) Topological protection ensures that the surface states'  conductivity $\sigma$ always increases with sample size $L$, as verified here near the resonance center $U = 7.2$,     off-center $U=9.1$, and in the large-disorder shoulder  $U=15.9, 30.0$. At very small $L$ tunneling through the bulk causes $\sigma$ to decrease with $L$. The variation in line slope with $U$ and with Fermi energy $E_F$ breaks the universality predicted by scaling theory.  
   (d) The $\beta$ function.  Scaling theory predicts a universal  $\beta = 2 / \pi \sigma $ curve independent of $U$ and $E_F$, shown here as a  black dotted-dashed  curve, and also shown in pane (a).  The deviations from universality seen here imply similar behavior in the magnetoconductivity.  
    }
\label{Fig3Conductivity}
\end{figure}

\section{The Conductivity}
In Figure 3 we turn to studying the topological metal's second signature, a conductivity $\sigma $ which grows robustly with sample size regardless of disorder strength.  We have carefully controlled for many effects and errors.   $\sigma $  grows only in the diffusive regime where several channels remain conducting, and here it is independent of sample width $W$.     Figure 3a  shows that both $\sigma $ and its  logarithmic derivative  $\beta(L) = d \ln \sigma / d \ln L $   converge to  their diffusive values
when $W > 1.1 \, L$; we restrict our remaining data to this converged regime. Moreover  in the diffusive regime both large slabs and large wires have the same conductivity; the gap is erased by disorder, and has no effect even at $E=0$. (See the supplemental material.)  Here we report  results obtained from slabs.    We also ensure convergence with slab  height $H$ by using thick slabs with $H=6$ in Figure 3a and $H=12$ elsewhere.  The associated computational cost is compounded by $\beta$'s sensitivity to statistical errors;  very large $N=960-4800$ numbers of samples - and smoothing in panes (a) and (c) -  were necessary to obtain these low-noise $\beta$ curves.   Leads effects are minimized by doping them into the metallic bulk band at $E_F=2$.

The most prominent feature of our data is  highlighted in  Figure 3c: a resonance  between the disordered surface states and the bulk band, seen here as a valley  in both $2/\pi\sigma$  and  $\beta$, which we have plotted  as  functions of disorder $U$  at four  Fermi levels inside the gap. It is centered around disorder strength $U=[6,10]$, matching the    bulk band width $\Delta E = 10$.   The resonance is generic to all TIs, since its physics is generic: surface disorder displaces the surface states into the bulk, as we already saw in Figure 2.   At weak disorder $U$ the displacement is small ($\delta = 0.06$ at $U=2$ in our model), but as $U$ passes through the resonance center the surface states migrate from the disordered surface layer into the clean bulk.  At the resonance center scattering is maximized, as is mixing between the bulk and the surface states.  Since quantum scattering processes are responsible for the  conductivity's growth, $\sigma$ is also maximized at the resonance center.  We conclude that outside the resonance center  the bulk tends to decrease the effect of surface disorder.  This will change the magnitude of the surface conduction in  TI samples and will also change the scattering length and the diffusion constant, each of which can be observed experimentally. 

   Figure 3c shows a very interesting feature:   at the resonance center  the four conductivity  curves   kiss,
   which signals that scattering  is independent of energy.  The surface density of states (DOS)  $\rho$ also must  be independent of energy, since it determines the scattering time  via $\tau \propto (\langle U^2 \rangle \, \rho)^{-1}$.  This is in remarkable contrast with the linear DOS $\rho(E) \propto E$  seen at zero disorder.

Figure 3b examines  the the conductivity growth signature of topological metals  at two values of the Fermi level $E_F=0,\,0.2$ and four representative disorder strengths.  The growth is very clear in the  two pink lines at the top which lie near the resonance center $U=7.2$, and also in the slightly lower two orange lines  which lie slightly off-center $U=9.1$.  The lowest four lines lying in the resonance shoulder $U=15.9, \,30.0$  do reveal a decreasing conductivity at small $L$, but this is a finite-size effect from the leads: disorder-assisted bulk tunneling  between the leads increases $\sigma$ in very short samples, and this excess decreases rapidly  with $L$. (See the supplemental material.) Leaving aside this tunneling effect, we find that $\sigma$ grows even when   its value (per surface) is as small as  $\sigma \geq 0.31 \frac{e^2}{h} \approx \frac{1}{\pi} \frac{e^2}{h}$.   This contrasts with materials without topological protection where any value of $\sigma$ up to $\sigma_C \approx 1.4 \frac{e^2}{h}$ produces a decreasing conductivity  \cite{Kawarabashi96,Kawabata03,Asada04,Asada06,Markos06,Nomura07}, and proves that  a TI's conductivity growth is robust against bulk effects.

The details of Figure 3 can be  compared with the one parameter scaling theory of conduction, which makes specific predictions about the diffusive regime.  Scaling theory's most important prediction is universality: the only effect of changing the disorder strength  and Fermi level should be to rescale both  the scattering length $l$ and  the overall length scale.  \cite{GangOfFour} The $\beta$ function is not sensitive to $l$, so it should be universal.  Numerical works on topological metals have shown that this universal curve  agrees quite well with   $\beta(\sigma) = 1/ \pi \sigma$,  even when $\sigma \approx 1/\pi$ is quite small.  \cite{Hikami80, Bardarson07,Nomura07,Mucciolo10}   
In consequence  $\sigma$ grows logarithmically $\propto 1/\pi \; \ln L$.   In our TI slabs $\beta$ and $\sigma$ are multiplied by $2$ for the two surfaces.
 In summary,  scaling theory predicts that in Figure 3d  the $\beta(\sigma)$ curves  should all coincide with each other and with   the $2/ \pi \sigma$ black dotted line, and that in Figure 3c each solid $\beta(\sigma)$ line should coincide with its partnering dashed $2 / \pi \sigma$ line.  Moreover 
   the conductivity curves in Figure 3b should all follow straight lines with the same slope $2/\pi$.  These universal results are at the origin of  the Hikami-Larkin-Nagaoka formula which gives a universal prediction for the conductivity's response to a small magnetic field, and in particular the $2 / \pi$ coefficient in these scaling theory predictions transfers over directly to the HLN formula's magnitude. \cite{Hikami80}

Near the resonance center $U\approx7.2$ we find excellent agreement with scaling theory, as evidenced by the pink straight line conductivity curves found in Figure 3b and by the pink  $\beta$ curves   in Figure 3d which coincide nicely with $2/\pi \sigma$.    The excellent agreement with scaling theory
 indicates that at the resonance center conduction is completely determined by diffusion and its quantum  corrections, and that the scattering length is very small. 

At other disorder strengths we find that scaling theory's universality is systematically violated.  We begin well within the resonance at $U=9.1$, which  is shown in the orange lines in Figures 3b,d.    These lines are straight, indicating that the conductivity grows with the dimensionless quantity $\ln L$, and proving that $\sigma$ is not controlled by any finite size effect.   However the lines' slope is clearly smaller than the $U=7.2$ slope (20\% smaller at E=0 and 12\%  at E=0.2),
 so $\beta$   is smaller than scaling theory's $2 / \pi \sigma$.   Figure 3c confirms this, showing that  at disorder strengths   near the resonance center the solid  $\beta$ lines lie consistently below the dashed  $2 / \pi \sigma$ lines.  This cannot be attributed to finite size effects or other errors, and is  an unambiguous signal  of non-diffusive conduction.

Turning to the resonance shoulder ($U=15.9, \;30.0$), at large $L$  we find that $\beta$ consistently exceeds the universal scaling theory prediction $2/\pi \sigma$ for  $E = 0.2, 0.3$,  as seen both  in Figure 3c and in  the blue dotted $E=0.2, W=30$ curves in Figure 3b,d.   This non-universal conduction   is likely superdiffusive, somewhere between diffusion and ballistic motion.  Once again this cannot be a finite size effect, since Figure 3b shows that in long $L=35$ samples $\sigma$ always becomes  roughly linear, i.e. proportional to $\ln L$.   This is confirmed by Figure 3d, which shows   $\beta$ converging toward  the decreasing $\beta \propto 1 / \sigma$ form which accompanies $\sigma \propto \ln L$.   We have checked that in longer $L\leq70$ samples all of the $\beta$ curves shown here begin decreasing.   
 We  conclude that the TI bulk reduces the topological metal's scattering and leaves the conductivity  in a non-diffusive, non-universal regime that is sensitive to sample details such as disorder strength and Fermi level.    This has immediate consequences for experimental measurements of the magnetoconductivity.  In particular,  its magnitude  should be sensitive to variations of the disorder strength and the Fermi level, in synchrony with the changing  magnitude of $\beta$ and $\sigma$.
 
section{Conclusions}
In summary, our results confirm that  the topological metal is robust against bulk effects when a finely-tuned magnetic field is applied, but also reveal that its response to disorder and to the magnetic field  is substantially changed by the bulk.  The bulk alters the PCC's decay, protects the topological metal from surface disorder except when the disorder is in resonance with the bulk, and pushes conduction into a non-diffusive, non-universal regime where it is sensitive to the Fermi energy and the disorder strength.

 \begin{acknowledgements}We acknowledge useful discussions with T. Ohtsuki, K. Kobayashi, B. A. Bernevig, D. Culcer, and A. Petrovic, and thank the Zhejiang Institute of Modern Physics and Xin Wan for hosting a workshop which allowed fruitful discussions.  This work was supported by the National Science Foundation of China and by the 973 program of China  under Contract No. 2011CBA00108. We thank Xi Dai and the IOP, which hosted and supported  this work.\end{acknowledgements}

  \begin{appendix}
  \section{Convergence of the PCC's Decay Length $\lambda$ in Slabs}

In Figure 4c in the main article we plot the PCC decay length $\lambda$ in slabs of width $W=3$.  The small width  allowed us to calculate very long slabs with lengths up to $L=10^6$.  Here we show that in slabs the decay length is roughly independent of slab width and is already close to convergence at $W=3$.  

Figure 4 here shows   $\lambda$ as a function of $W$  in slabs with cross-section $H \times W$.  The left-most data points at $W=3$ are shown in the main article's Figure 4c.  The disorder strength is $U=2$, the Fermi energy in both the leads and the sample is $E_F = 0.7$, and the number of samples is bounded below by $N \geq 240$. 

The only significant convergence problem is an oscillation at small $W$ due to state quantization across the width of the slab.   $\lambda$'s value at $W=3$ is pretty close to its value at $W=40$.  Moreover the difference between $\lambda$'s values at  different heights $H$ is independent of  $W$.  We conclude that the width $W=3$ used in the main article causes only small errors in our results.

\section{Determination of the Scattering Length}
        \begin{figure}[]
\centering
\includegraphics[scale=0.33,clip]{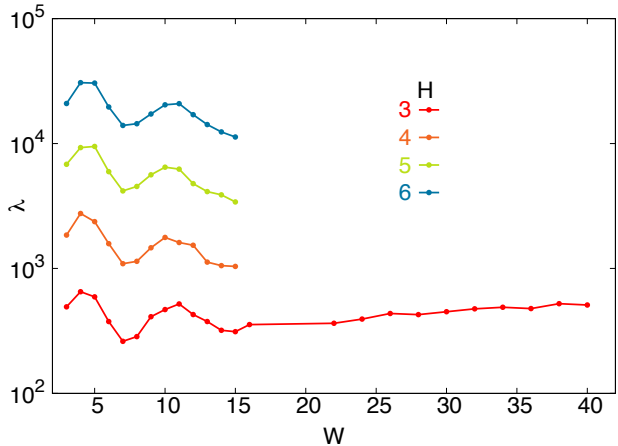}
\caption{ (Color online.) Convergence of the PCC decay length $\lambda$ with slab width $W$.  Curves at four slab heights  $H=3,4,5,6$ are equally spaced, showing that  the linear coefficient of $\ln \lambda$'s dependence on $H$ is independent of $W$.     }
\label{Fig1PCC}
\end{figure}

        \begin{figure}[]
           \centering
\includegraphics[angle=90,scale=0.65,clip]{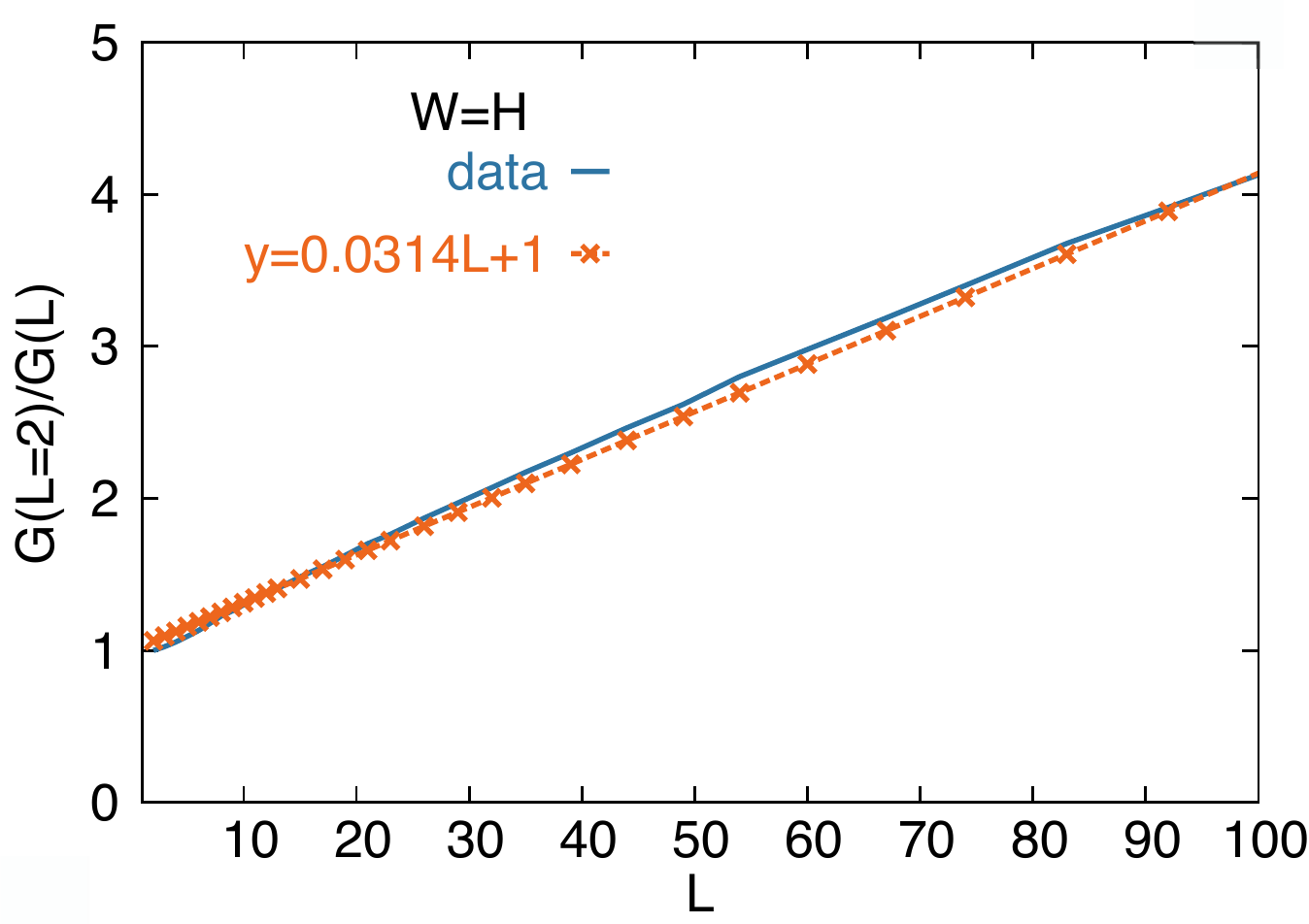}
\caption{ (Color online.)  Scattering length at disorder strength $U=2$. We show $G(L=2)/G(L)$ as a function of wire length $L$ in a $10 \times 10$ wire. The dotted line shows a best linear fit.  Its slope, $0.031$, is the inverse scattering length $\lambda^{-1}$.   $E_F = 0.7$ in both the sample and the leads, and the disorder strength is $U=2$. The number of samples is $N=864$.}
\label{Fig2RShiftG}
\end{figure}
Figure 5 supports our measurement of the scattering length $\lambda \approx 30$, which we obtain by fitting the conductance to  $\frac{G(L=0)}{G(L)}  = (L/\lambda) + \rho$, where $\rho$ includes the physics of the contact resistance. \cite{Beenakker97} We use $G(L=2)$ instead of $G(L=0)$, which causes only a small error in our fitting.   The scattering length is determined to be $\lambda = (0.031)^{-1}$. 

\section{Large Slabs and Wires Give Identical Results in the Diffusive Regime}
In the main article we state that in the diffusive regime both large slabs and wires have the same conductivity.  Here in Figure 5 we plot raw $\sigma$ data for both types of samples, as a function of  finite wire circumference and finite wire width.  We set the disorder strength $U=7.2$ in the resonance, and the sample height is $H=6$ for both wires and slabs.  We set the Fermi energy $E_F= 0.0$ in the sample and $E_F = 2.0$ in the leads, but identical results are obtained when $E_F = 0.2$ in the sample.  The number of statistics is $N=12000$ at $W=5,10$, $N=9600$ at $W=15$, $N=4800$ at $W=20$, $N=2400$ at $W=30$, $N=960$ at $W=40$, and  $N=480$ at $W\geq50$.  

The main result of these figures is that the conductance $\sigma$ converges quickly with sample width/circumference, and that at $W>40$ there is little or no difference between slabs and wires.  In particular, the Berry phase gap has little effect at $W>40$ for two reasons: (1) The gap size scales with $1/W$, and (2) the sample is in diffusive regime, where the disorder  broadens the band edges and erases the gap.

      \begin{figure}[]
           \centering
 \includegraphics[angle=90,scale=0.45,clip]{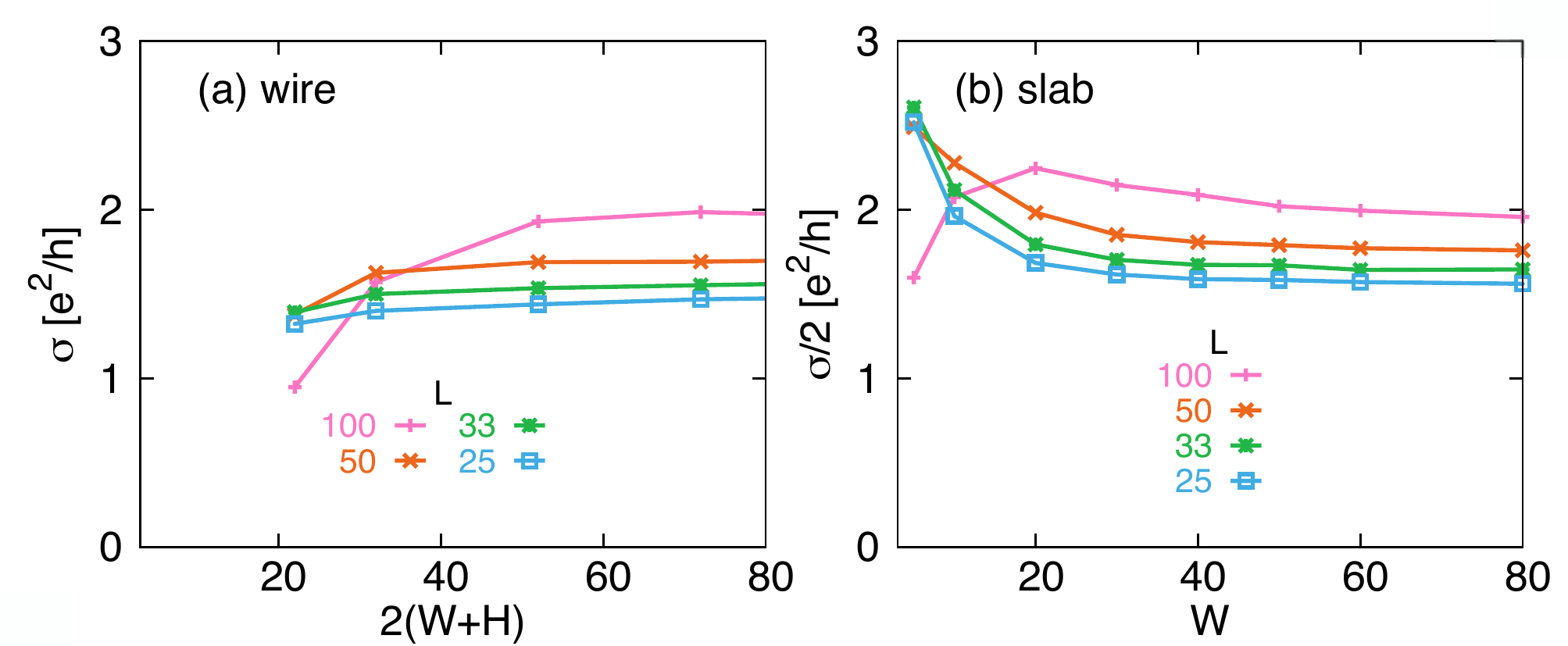}
\caption{(Color online.) Large diffusive slabs and wires give identical conductivities.  (a)  Convergence with wire circumference $2(W+H)$ of $\sigma$ in wires.   (b)  Convergence with slab width $W$ of $\sigma$ in slabs with lengths $L=25,\,33, \,50,\,100$. The disorder strength is in the resonance $U=7.2$ and the Fermi energy energy is $E_F=0$.  Identical results are found at $E_F=0.2$. }
\label{Fig2RShiftB}
\end{figure}

\section{Disorder-Assisted Tunneling Between the Leads in Very Short Slabs}
           \begin{figure}[]
           \centering
\hspace*{-3cm} \includegraphics[scale=0.55,clip]{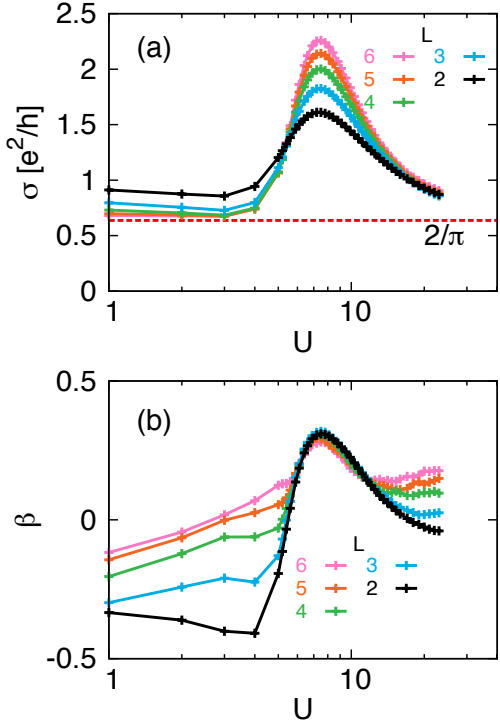}
\caption{ (Color online.) Disorder-assisted tunneling in very short TI slabs.  We plot $\sigma(L)$ and its derivative $\beta(L)$ as functions of disorder strength $U$ in slabs with  very short lengths $L=2,3,4,5,6$.  }
\label{Fig2RShiftC}
\end{figure}
In the main article we briefly discuss disorder-assisted tunneling between the leads, which in short samples increases the conductivity and reverses the sign of the beta function and its derivative.  
Here in Figure 7 we highlight disorder-assisted tunneling between the leads, which occurs in very short samples.  This is raw data obtained by averaging $N=4320$ samples.  We use slabs with a $6 \times 60$ cross-section, and set the Fermi energy at $E=0$ in the sample and $E=2$ in the leads.   

The surface states'  resonance with the bulk band is visible as a peak in  both $\sigma$ and $\beta$. Inside this peak the derivative $\beta(L)$ is a positive and slowly decreasing function of $L$, as expected from scaling theory.    On the wings of the resonance $\beta$ is a negative and increasing function of $L$, because  tunneling between the leads increases the conductance.  This tunneling has the greatest effect at $L=2$, and decays rapidly with sample length, causing the decreasing conductivity and negative beta.  
 
This data should be  compared to  wide $W \gg L$ graphene strips, where at small $L$ the conductivity converges to $\frac{4}{\pi}\frac{e^2}{h}$, and at other $L$ the conductivity is always greater than $\frac{4}{\pi}\frac{e^2}{h}$. \cite{Mucciolo10}  Moreover at small $L$ graphene's $\beta$ converges to zero.  Like graphene, our data  does show that in TIs at the Dirac point the conductivity is never smaller than $\frac{2}{\pi}\frac{e^2}{h}$.  (The factor of $2$ difference is associated with the different number of Dirac cones in TIs.)   However, unlike graphene our data  also shows that in TIs the conductivity at $L\rightarrow0$ depends on disorder and is always larger than  $ \frac{2}{\pi}\frac{e^2}{h}$, presumably because disorder increases the coupling between the leads.  This is manifested also in the $L \rightarrow 0$ limit of $\beta$, which is nonzero and depends on disorder.

\section{Convergence of the Conductivity and the $\beta$ Function with Slab Height and Width}
           \begin{figure}[]
          \centering
           \includegraphics[angle=90,scale=0.47]{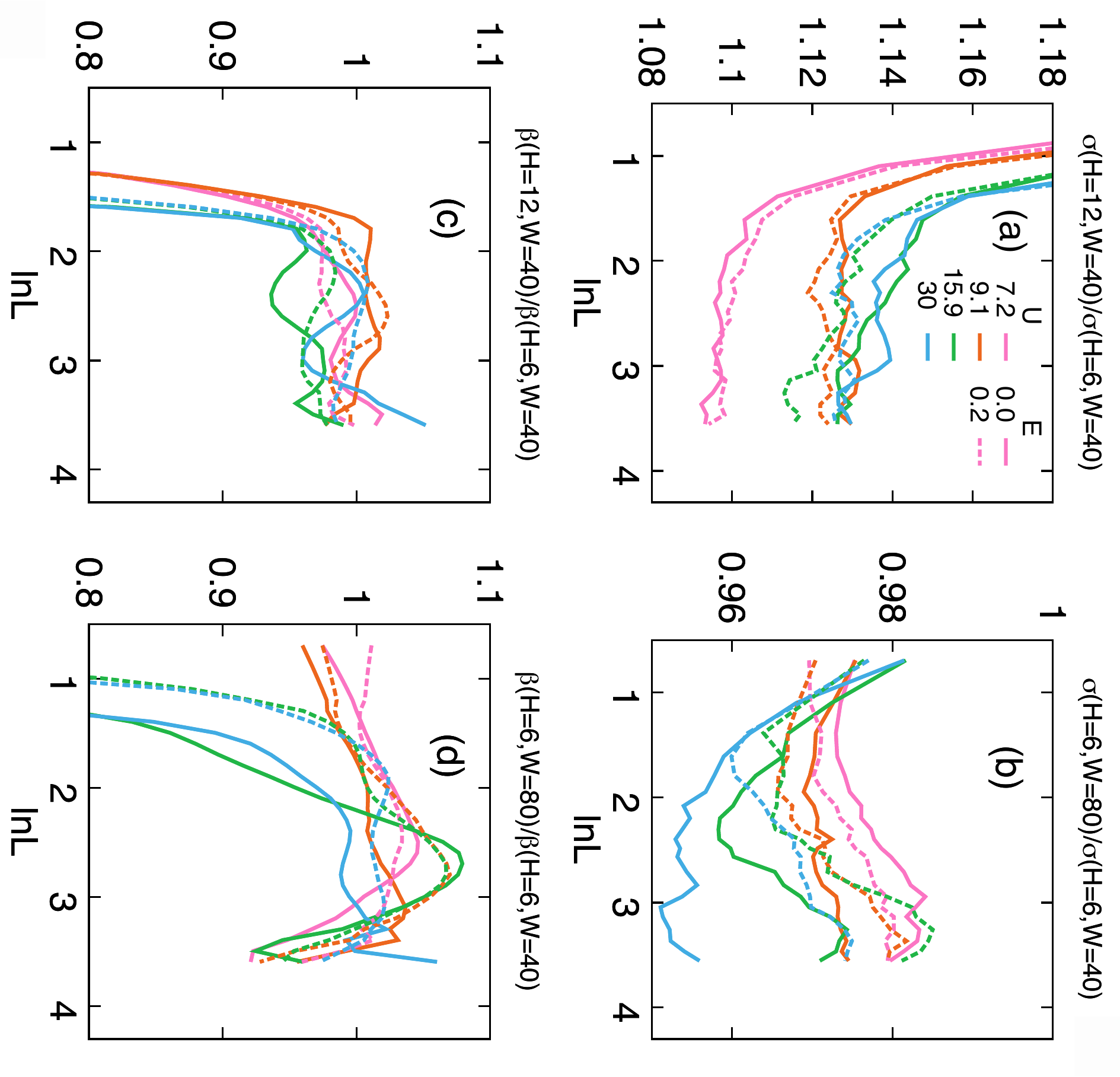}
\caption{ (Color online.) Finite size effects on the conductivity $\sigma$ and on the $\beta$ function.  (a) Corrections to the conductivity caused by the slab height $H$.  We compare $H=6$ to $H=12$ and keep the width fixed at $W=40$. (c) Corrections to $\beta$ caused by the slab height $H=6,12$, with $W=40$.  (b) Corrections to the conductivity caused by slab width $W$.  We compare $W=40$ to $W=80$ and keep the height fixed at $H=6$.  (d) Corrections to $\beta$ caused by slab width $W=40,80$ with $H=6$.   We used $N=4800$ samples for the $H=12, W=40$ data, $N=8640$ samples for the the $H=6,W=40$ data, and $N=2160$ samples for the $H=6, W=80$ data. }
\label{Fig2RShiftD}
\end{figure}

Here in Figure 8 we examine finite size corrections to the $\sigma$ and $\beta$ curves in panes b and d of the main article's Figure 8.  We use the same parameters as in the main article's Figure 3.  The only difference is that in panes a and c of Figure 8 we change the slab height from $H=12$ to $H=6$, and then plot the ratios of $\sigma$'s and $\beta$'s values at these  heights.  Similarly in panes b and d of Figure 8  we change the slab width from $W=40$ to $W=80$ while keeping the slab height fixed at $H=6$, and then plot the ratios of $\sigma$'s and $\beta$'s values at these two widths $W=40,80$.  In order to reduce noise in $\beta$ we have splined and smoothed the conductivity data prior to computing $\beta$.

Panes a and c of Figure 8 compare the  slab heights $H=6,12$.  We expect that $H=12$ is quite close to the $H=\infty$ limit because height effects should be regulated by an exponential.  At $L > 7$ the main effect of tunneling in thin slabs is to multiply the conductivity by a factor which is less than one.    Figure 8a show that this factor  is very weakly dependent of length $L>7$, and is roughly independent of the Fermi energy $E$.  However the tunneling  factor does depend on the disorder strength $U$ - it is a 10\% effect at the resonance center $U=7.2$, and a 13-14\% effect at other disorder strengths.   Because the tunneling factor depends only weakly on length $L>7$, the tunneling effect on $\beta$ is rather small, as seen in pane (c).  Since the $H=6$ tunneling errors in both $\beta$ and $\sigma$ are small, we expect that tunneling errors are negligible in the $H=12$ data presented in the main article's Figure 3.

Panes b and d of Figure 8 compare the  slab widths $W=40,80$.  This error should be regulated by $1/W$, so the difference between $W=80$ and $W=\infty$ will be of the same magnitude as the difference plotted here between $W=40$ and $W=80$.  Pane (b) shows that in smaller slabs the conductivity is overestimated by a few percent. We have considered correcting the $W=40$ data in the main article's Figure 8 to account for the overestimate of $\sigma$, but the correction makes little visible impact on Figure 8 and no change in our conclusions.     Pane (d) shows that in $W=40$ slabs at $L>7$ the finite size error in the $\beta$ function is small enough that it is difficult to distinguish from statistical noise, and depends in a complicated way on the Fermi level, the disorder strength, and the sample length.  $\beta$ may be either overestimated or underestimated by as much as 7\%, but usually around 3-4\%, close to the noise level.  Both the complicated profile and the small magnitude of the finite size error in $\beta$  imply that this error does not affect our conclusions.

\section{Convergence of the Conductance with Height and Width in Slabs and Wires}
           \begin{figure}[]
           \centering
\includegraphics[angle=90,scale=0.43]{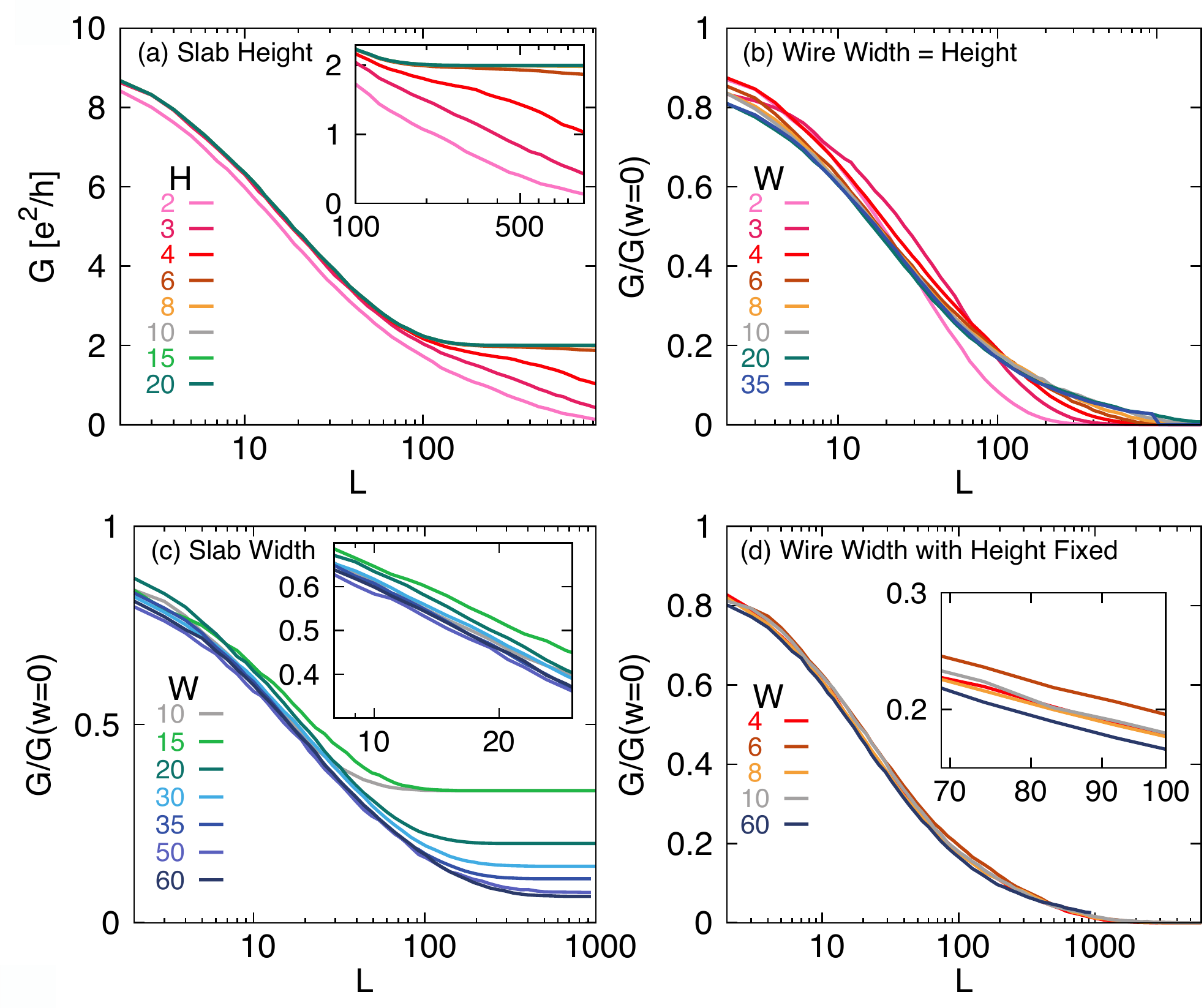}
\caption{ (Color online.) Convergence of the conductance $G$ as a function of cross-section in short wires and slabs. (a) The conductance as a function of slab  length $L$, at many slab heights $H$.   The inset magnifies the $L=[100,800]$ data. The slab width is $W=20$.  Panes (b-d) show the conductance at disorder $U=2$, divided by its quantized value in pure samples where the conductance is independent of $L$. (c) Convergence with slab width $W$, with a closeup shown in the inset. (b) Convergence with $W=H$ in square wires.  (d) Convergence with width $W$ in rectangular wires of height $H=16$, with a closeup shown in the inset.  }
\label{Fig2RShiftE}
\end{figure}

Panes 3b,3c, and 3d of the main article's Figure 3 plot the conductivity and beta function in slabs of width $W=44$ and height $H=12$.  Here in Figure 9 we examine  conductance's convergence as a function of cross-section in both slabs and wires.

  In pane 9a we keep the slab width fixed at $W=20$, so that the conductance should converge to a height-independent value.  In panes 9b-9d we obtain a convergent observable by dividing the conductance at disorder strength $U=2$ by its value in a pure sample where the conductance is independent of length.   In all panes the disorder strength is $U=2$, and the Fermi energy is $E_F = 0.7 $ both in the leads and in the sample.  The number of samples is $N=240$ in panes 9a, 9c, and 9d.  In Pane 9b we calculated $9600$ wires with  $2 \times 2$  and $3 \times 3$ cross-sections, $4800$ wires with a $4\times4$ cross-section,   $480$ wires with  $6\times 6$ and $8 \times 8$ cross-sections, $120$ wires with  $10 \times 10$ and $20 \times 20$ cross-sections, and finally $60$ wires with a  $35 \times 35$ cross-section.

Pane 9a shows that in slabs  $H=3$ is sufficient to obtain good convergence at $L<100$ and that $H=6$ gives good convergence out to $L=800$.  Pane 9c shows that in $H=16$ slabs fairly good convergence is obtained  at $W=20$ but $W=40$ gives even better results.  Interestingly, $W=15$ gives poorer results than $W=10$. The delayed convergence at $L\geq 80$  is caused by the  PCCs which cause $G(U=2)/G(L=0)$ to scale with $1/W$. Pane 9b shows that in $W \times W$ wires $W=6$ already gives good convergence at $L < 300$.  Pane 9d shows that in $W\times 16$ wires fairly good convergence is obtained at $W=4,8,10$, but the best convergence is found when $W>40$.

From Figure 9a, in conjunction with Figure 4 in the supporting material, we conclude that the surface states' penetration depth is very short, and that convergence with respect to slab height is controlled by an exponential.  As long as $L < 100$ we can expect very thin $H=3,4$ slabs to give nearly converged results.  These results were obtained at disorder strength $U=2$, and we can expect some change at larger disorders $U=8, 40$. However since we are using surface disorder  we expect the  convergence threshold in slabs to shift by only $2$, to  $H=5,6$.  The small-$W$ data in figures 9b and 9d show the same very  fast decay of tunneling in thin wires.  

Figures 9b and 9d also show that convergence is not fully achieved until dimensions of $30-60$ are reached.  This is not a tunneling effect, but instead a finite size effect caused by the finite lateral dimension in slabs, and by the finite circumference in wires.  It is small compared to the tunneling effect, but persists to much larger cross-sections.  

In conclusion, the results of Figure 9 inform us that in the main article's Figure 3  tunneling effects should be small  because the height is $H=12$, and finite size effects should also be small because the width is $W=44$.

\section{Agreement of $\beta$'s First Derivative with Scaling Theory}
      \begin{figure}[]
  \includegraphics[angle=90,scale=0.65]{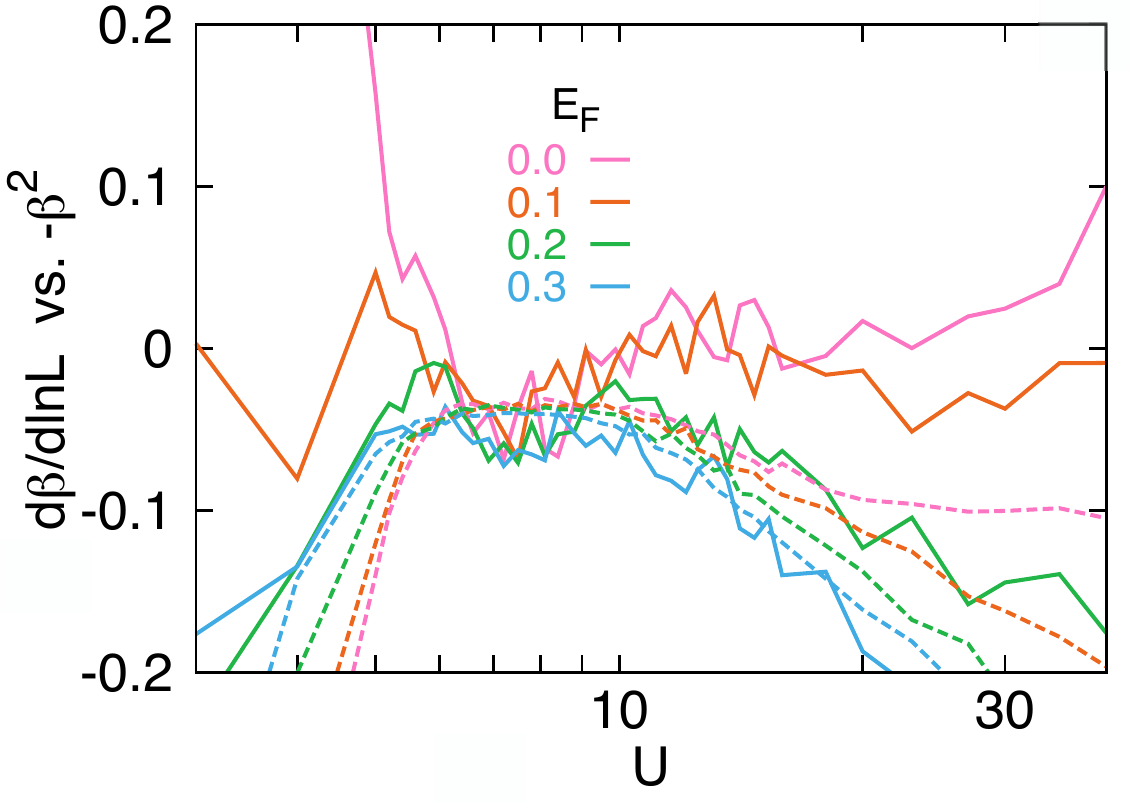}
\caption{ (Color online.) $\beta$'s first derivative.   $\frac{d\beta(L)}{d \ln L} $ (solid lines) vs. the scaling theory prediction $-\beta^2$ (dashed lines) at $L=26$ and four Fermi levels.        }
\label{Fig2RShiftF}
\end{figure}

In the main article's Figure 3c we checked the agreement of  $\beta$ with the scaling theory prediction $2/\pi \sigma$, as a function of disorder strength $U$ and at four Fermi levels. Here  in Figure 10 we examine  $\beta$'s first derivative $\frac{d\beta(L)}{d \ln L}$, which should  be equal to $-\beta^2$ if scaling theory holds.  Both $\beta$ and its first derivative were calculated at $L=26$  after doing a linear fit to our raw $\beta$ data  over the range $\ln L=[\ln 5, \ln 54]$.  We used $N=480$ $6\times 60\times 60$ slabs for this calculation.  Our results show that in the resonance center the scaling theory prediction is roughly verified, but outside of the center the agreement vanishes.  The main article's Figure 3d shows that when we  increase the slab size  and the length $L$ the disagreement seen here is considerably diminished at large $L$, hinting that $\frac{d\beta(L)}{d \ln L}$'s agreement with  $-\beta^2$ also may  improve at large $L$.

\end{appendix}

\bibliography{Vincent}
\end{document}